\documentclass[a4paper,11pt]{article}
\usepackage{pos}
\usepackage{gensymb}
\usepackage{caption}
\usepackage{float}

\title{Atmospheric muon suppression for Baikal-GVD cascade analysis}
\ShortTitle{Suppression of Background Cascades}

\author[a]{V.M.~Aynutdinov}
\author[b]{V.A.~Allakhverdyan}
\author[a]{A.D.~Avrorin}
\author[a]{A.V.~Avrorin}
\author*[c,d]{Z.~Barda\v{c}ov\'{a}}
\author[b]{I.A.~Belolaptikov}
\author[a]{E.A.~Bondarev}
\author[b]{I.V.~Borina}
\author[e]{N.M.~Budnev}
\author[l]{V.A.~Chadymov}
\author[f]{A.S.~Chepurnov}
\author[b,g]{V.Y.~Dik}
\author[a]{G.V.~Domogatsky}
\author[a]{A.A.~Doroshenko}
\author[c]{R.~Dvornick\'{y}}
\author[e]{A.N.~Dyachok}
\author[a]{Zh.-A.M.~Dzhilkibaev}
\author[c,d]{E.~Eckerov\'{a}}
\author[b]{T.V.~Elzhov}
\author[d]{L.~Fajt}
\author[l]{V.N. Fomin}
\author[e]{A.R.~Gafarov}
\author[a]{K.V.~Golubkov}
\author[b]{N.S.~Gorshkov}
\author[e]{T.I.~Gress}
\author[h]{K.G.~Kebkal}
\author[a]{I.V.~Kharuk}
\author[b]{E.V.~Khramov}
\author[b]{M.M.~Kolbin}
\author[i]{S.O.~Koligaev}
\author[b]{K.V.~Konischev}
\author[b]{A.V.~Korobchenko}
\author[a]{A.P.~Koshechkin}
\author[f]{V.A.~Kozhin}
\author[b]{M.V.~Kruglov}
\author[j]{V.F.~Kulepov}
\author[e]{Y.E.~Lemeshev}
\author[a,\dagger]{M.B.~Milenin}
\author[e]{R.R.~Mirgazov}
\author[b]{D.V.~Naumov}
\author[f]{A.S.~Nikolaev}
\author[a]{D.P.~Petukhov}
\author[b]{E.N.~Pliskovsky}
\author[k]{M.I.~Rozanov}
\author[e]{E.V.~Ryabov}
\author[a]{G.B.~Safronov}
\author[b,g]{D.~Seitova}
\author[b]{B.A.~Shaybonov}
\author[a]{M.D.~Shelepov}
\author[a]{S.D.~Shilkin}
\author[f]{E.V.~Shirokov}
\author[c,d]{F.~\v{S}imkovic}
\author[b]{A.E.~Sirenko}
\author[f]{A.V.~Skurikhin}
\author[b]{A.G.~Solovjev}
\author[b]{M.N.~Sorokovikov}
\author[d]{I.~\v{S}tekl}
\author[a]{A.P.~Stromakov}
\author[a]{O.V.~Suvorova}
\author[e]{V.A.~Tabolenko}
\author[b]{B.B.~Ulzutuev}
\author[b]{Y.V.~Yablokova}
\author[a]{D.N.~Zaborov}
\author[b]{S.I.~Zavyalov}
\author[b]{D.Y.~Zvezdov}

\affiliation[a]{Institute for Nuclear Research, Russian Academy of Sciences, Moscow, 117312, Russia}
\affiliation[b]{Joint Institute for Nuclear Research, Dubna, 141980, Russia}
\affiliation[c]{Comenius University, 81499 Bratislava, Slovakia}
\affiliation[d]{Czech Technical University in Prague, Institute of Experimental and Applied Physics, 11000 Prague, Czech Republic}
\affiliation[e]{Irkutsk State University, Irkutsk, 664003, Russia}
\affiliation[f]{Skobeltsyn Institute of Nuclear Physics, Moscow State University, Moscow, 119991, Russia}
\affiliation[g]{Institute of Nuclear Physics of the Ministry of Energy of the Republic of Kazakhstan, Almaty, 050032, Kazakhstan}
\affiliation[h]{LATENA, St. Petersburg, 199106, Russia}
\affiliation[i]{INFRAD, Dubna, 141981, Russia}
\affiliation[j]{Nizhny Novgorod State Technical University, Nizhny Novgorod, 603950, Russia}
\affiliation[k]{St.~Petersburg State Marine Technical University, St.~Petersburg, 190008, Russia}
\affiliation[l]{Moscow, free researcher}

\note[\textdagger]{Deceased.}

\emailAdd{zuzana.bardacova@fmph.uniba.sk}

\abstract{Baikal-GVD (Gigaton Volume Detector) is a neutrino telescope installed at a depth of 1366 m in Lake  	Baikal.   The expedition of 2023 brought the number of optical modules in the array up to 3492 (including experimental strings). These optical modules detect the Cherenkov radiation from secondary charged particles coming from the neutrino interactions. Neutrinos produce different kinds of topologically distinct light signatures. 
	Charged current muon neutrino interactions create an elongated track in the water. 
	Charged and neutral current interactions of other neutrino flavors yield  hadronic and electromagnetic cascades.  
	 The background in the neutrino cascade channel arises mainly due to discrete stochastic energy losses produced along atmospheric muon tracks. 
	 In this paper, a developed algorithm for the cascade event selection  is presented.}

\FullConference{%
	38th International Cosmic Ray Conference (ICRC2023)\\
	26 July - 3 August, 2023\\
	Nagoya, Japan}
\ConferenceLogo{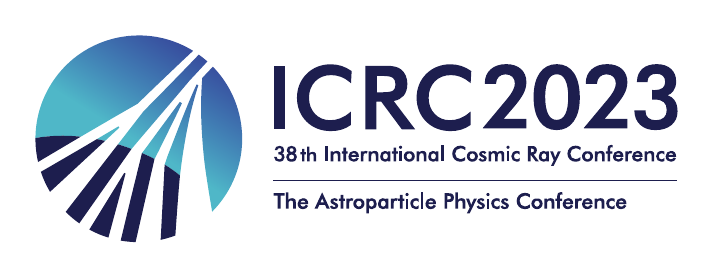}

\begin{document}
	\maketitle

	\section{Introduction}
	Baikal-GVD (Gigaton Volume Detector)  is a cubic kilometer scale neutrino observatory located in the southern part of Lake Baikal, Siberia. Currently (year 2023), its 3492 light sensors (optical modules - OMs, including experimental strings) detect the Cherenkov light that is produced by secondary charged particles originating from neutrinos interacting in the Baikal water. A three-dimensional array of OMs   organized in so-called clusters is located at a depth of 1366m, about 3.6 km offshore \cite{GVD}. The primary purpose of Baikal-GVD is to search for high-energy neutrinos that originate from the same cosmic particle accelerators, which produce very high energy cosmic rays. Moreover, the diffuse
	flux emitted collectively by  unresolved astrophysical sources can be observed \cite{diffuse}. 
		
\begin{figure}[h!]
		\centering
		\includegraphics[scale=0.26]{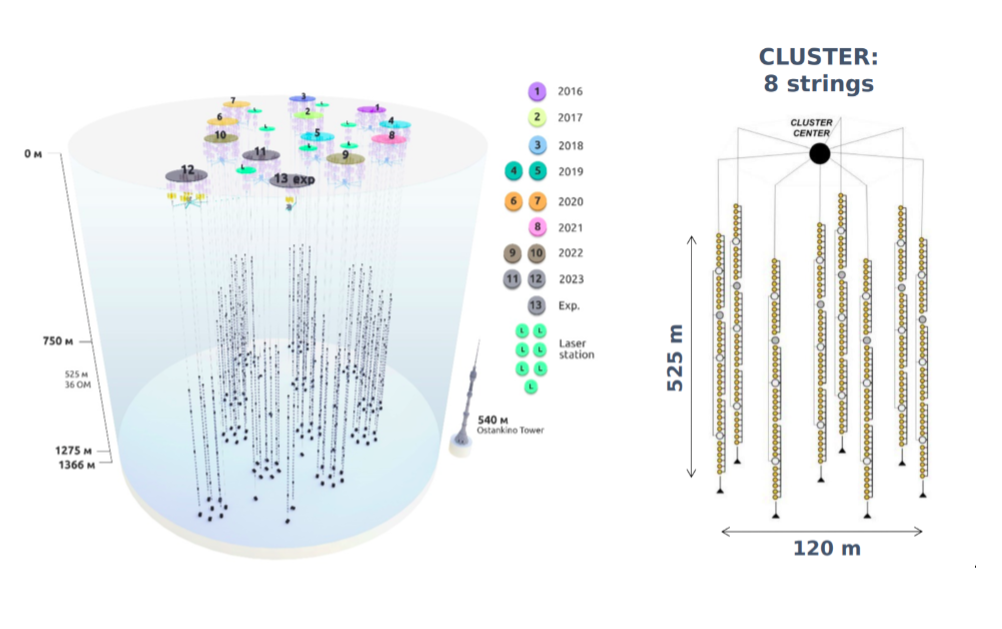}
		\caption{Left: Baikal-GVD in 2023. The detector is composed of individual clusters, laser stations and experimental strings. The cluster color scheme represents annual deployment progress. Right: A standard Baikal-GVD  cluster with 8 strings.}
		\label{GVDsketch}
	\end{figure}

		Thirty-six OMs are installed on each of the 96 vertical strings with the distance of 15 m between adjacent OMs, from 750 m to 1275 m below the surface. Most of the strings are arranged into independently operated clusters, while each cluster is composed of 8 strings. Each OM contains a 10-inch photo-multiplier tube (PMT) Hamamatsu R7081-100 housed in a 13-inch glass sphere.
 The schematic view of the Baikal-GVD detector is shown in Fig. \ref{GVDsketch}.
	
	Baikal-GVD  primarily observes two  topologically distinct classes of events: tracks and cascades. The charged current interaction of a muon neutrino ($\nu_{\mu}$) with matter results in an outgoing muon, which travels a long distance in water and leaves an elongated track signature. The cascade events arise from the interactions of all three neutrino flavours. The charged current (CC) electron neutrino ($\nu_e$) interaction and the neutral current (NC) electron, muon, and tau neutrino ($\nu_e$, $\nu_{\mu}$, $\nu_{\tau}$) interactions produce detectable cascade light signatures. In case of  cascades most of the neutrino  energy is deposited in a small volume, that results in a nearly spherical event. An advantage of the cascade channel over a track channel is that cascade events typically have allow for a better energy resolution, because the events can be fully contained inside the detector.  However, it is more difficult in the case of cascades to reconstruct the initial direction of the neutrino.
	
	At the angles above the horizon, there is an overwhelming background of muons produced in air showers when cosmic rays enter the Earth's atmosphere. Muons from the air showers come in bundles containing up to hundreds of muons. In the cascade channel, the main background originates from the discrete stochastic  processes  along the muon track as a result of bremsstrahlung, photonuclear processes or direct electron-positron pair production (see Fig. \ref{sketch}, left). Conventional rejection strategy for the atmospheric muon bundles is to select only events coming from the lower part of the detector (upgoing). However, the background cascade events may be  wrongly reconstructed as upgoing, while they were truly downgoing muons.
A search for the signal cascades is, therefore, challenging due to the high muon flux  from the air showers.

In this work, developed and optimized techniques to separate  the neutrino-induced cascades from the background (cascades from atmospheric muon bundles) are discussed. The main difference
between signal and the background cascade is the presence of a muon track.  The Cherenkov light
from the muon track can change the cascade light signature and  influence the cascade reconstruction variables.  For development of the selection methods we used   Monte Carlo (MC) simulations for the part of the 2019 season from April to June (5 cluster configuration).  
Each selection method provides an output
variable, which are fed into a Boosted Decision Tree (BDT) \cite{BDT}. The selection algorithm was optimized only with single-cluster data.
Furthermore, the selected experimental neutrino cascade candidate was searched for among the multicluster events to find an indication of a muon track. Moreover, preliminary waveform analysis of that interesting event was performed.   
	
	\section{Background Cascades}
	\label{Sup}
\begin{figure}[h!]
	\centering
	\includegraphics[scale=0.27]{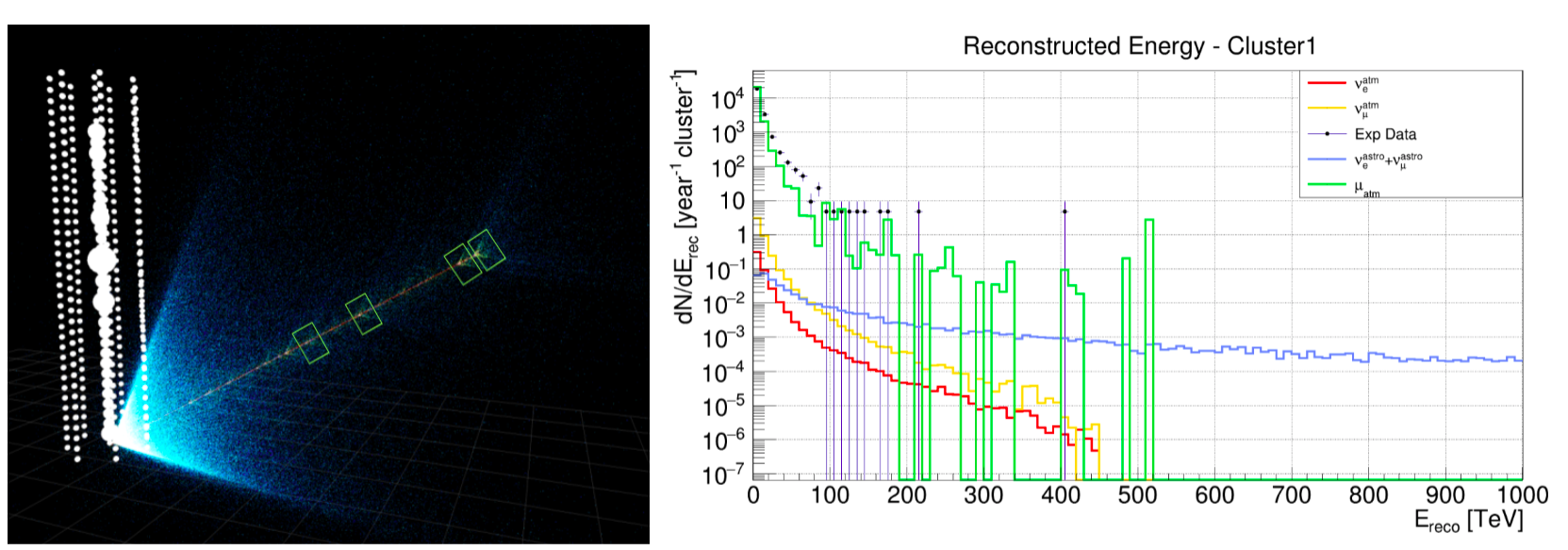}	
	\caption{Left: The characteristic topology of light emission for a muon track (blue light) with visible stochastic light depositions along the track (background cascades - marked by green boxes).	 Right: Reconstructed energies for cascade-like events. Black dots display the experimental data from Cluster 1 from season 2019. The predicted $\mu_{\rm{atm}}$ background is shown by the green line, the atmospheric neutrinos $\nu^{\rm{atm}}_e$  and $\nu^{\rm{atm}}_{\mu}$ are shown by the red and yellow lines, respectively.
		Astrophysical $\nu^{\rm{astro}}_e$ and $\nu^{\rm{astro}}_{\mu}$ neutrinos are merged into one dataset and shown by the light blue line.}
	\label{sketch}
\end{figure}

	In this work,  MC data sets for signal and background were generated for each cluster separately and used for development of the selection techniques. The arrangement of each cluster is set up to correspond to the average conditions during the initial phase of the 2019 season (between April 1 and June 30). During this interval, the optical activity (luminescence) of the lake is relatively low, and the noise rates fluctuate from $\approx$ 15 kHz (for the bottom OMs) to $\approx$ 50 kHz (for the OMs located at the uppermost sections) \cite{noise}.	For comparison of the experimental data and MC samples we used runs from the same period with the effective livetime of  353 days (combined for all 5 clusters).
	Atmospheric $\nu^{\rm{atm}}$  and astrophysical $\nu^{\rm{astro}}$ neutrino events (only electron and muon) were considered as a signal. In the MC signal simulation dataset, neutrino energies range from 1 TeV to 400 TeV and from 1 TeV to 400 PeV for atmospheric and astrophysical neutrinos, respectively.  The energy interval of cosmic ray protons in the background MC sample is from  240 GeV to 100 PeV. After MC simulations a reconstruction of the cascade energy and direction was used. A parallel package \cite{parallel} has been used for the cascade reconstruction software.
	Fig. \ref{sketch} (right)	displays the event rate for the experimental data and MC datasets as a function of the reconstructed cascade energy. The cut applied on the events was that the horizontal distance of the reconstructed vertex	position from the centre of the cluster can be no more than $\rho = 100$ m. The rate of events is shown per cluster per one year. It can be observed that most of the reconstructed cascade-like events from the experimental data follow the MC background cascades from $\mu_{\rm{atm}}$. According to the MC simulations the background cascades can also reach high energies. Note also that the event rate from downgoing $\mu_{\rm{atm}}$  is almost  4 orders 	of magnitude higher than the event rate from atmospheric neutrinos. Hence, the development of methods for the selection of neutrino cascades is essential.

\section{Neutrino Cascade Selection Algorithm} 
Various selection methods for neutrino cascade events in the Baikal-GVD were implemented, tested, and optimized. Described technique for the selection analysis of signal and background cascades represents an additional step in the cascade selection algorithm  \cite{ECRS2022}. The optimization was performed on the simulated data of background cascades from $\mu_{\rm{atm}}$ and signal cascades from   $\nu^{\rm{atm}}$  and $\nu^{\rm{astro}}$ interactions. For the analysis we only took into account  contained cascade-like events reconstructed as upgoing. 

\begin{figure}[h]
	\centering
	\includegraphics[scale=0.35]{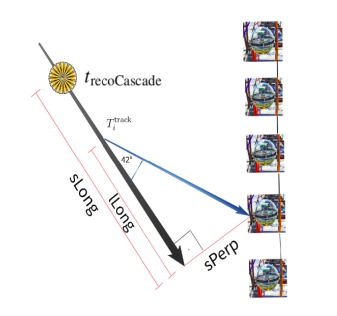}
	\includegraphics[scale=0.34]{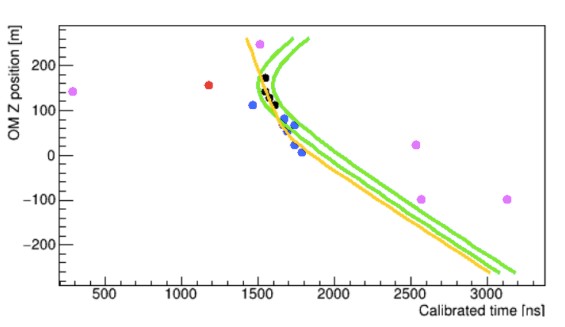}
	\caption{Left: Geometry components used for the calculation of the predicted OM hit time from the muon track in Eq. \ref{eqTrack}. Right: The OM z coordinate (on one string) as a function of the OM hit time for simulated $\mu_{\rm{atm}}$ event. Red dot corresponds to $t_{\rm{recoCascade}}$, track hits are shown in blue, cascade hits in black and noise hits in pink color. Yellow line corresponds to the predicted time $T^{\rm{track}}_i$ for track hits according to  Eq. \ref{eqTrack}. Green lines show the expected time interval for the cascade hits. Note that track hits are detected earlier with respect to $t_{\rm{recoCascade}}$  compared to the cascade hits.}
	\label{trackSketch}
	\end{figure}

\begin{figure}[h]
	\centering
	\includegraphics[scale=0.43]{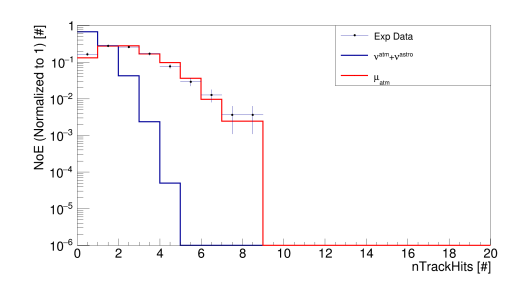}
	\includegraphics[scale=0.29]{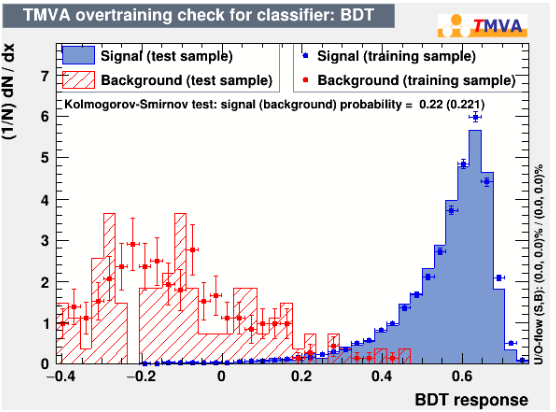}
	\caption{nTrackHits (left)  distribution for the signal cascades (blue line), background cascades (red line), and experimental data (black dots). BDT response score (right) for background cascades (red histogram) and signal cascades (blue).}
	\label{nTrackHits}
\end{figure}

First, we developed the nTrackHits method that differentiates background and signal cascades by leveraging the existence of a muon track within the atmospheric muon bundle and its absence in the neutrino cascade events.  The nTrackHits method selects hits  with  time residuals in the interval:
 
\begin{equation}
		t_{\rm{1}} < t_i - T^{\rm{track}}_i < t_{\rm{2}},
\end{equation}
	where $t_i$ is the OM hit time, $t_1 = -100 ~\rm{ns}$, $t_2 = 25$ ns. The expected time $T_i$, when the OM is supposed to detect  a hit from  the muon track is obtained from:
\begin{equation}
		T^{\rm{track}}_{i} = t_{\rm{recoCascade}} + \rm{(sLong - lLong)}\cdot \frac{1}{c} + \sqrt{\rm{sPerp}^2 + \rm{lLong}^2}\cdot \frac{1}{c_w},
		\label{eqTrack}
\end{equation}
	where $t_{\rm{recoCascade}}$ is time of the reconstructed cascade, $c_w$ is the speed of light in water,  and $c$ is the speed of light in vacuum (see Fig. \ref{trackSketch} (left)). The track direction cannot be determined as the reconstructed cascade direction, because it can be misreconstructed
	due to track hits that were incidentally used in the reconstruction. Therefore, the
	nTrackHits is determined in each muon direction given by the iteration of the reconstructed cascade direction over azimuth and	zenith angle in a cone with apex angle $\approx 40 \degree$.
	The OM hit time $t_i$ dependence on the OM z position for the reconstructed background cascade from simulated MC $\mu_{\rm{atm}}$ event is displayed in Fig. \ref{trackSketch} (right). Each color indicates a different origin of the hit. The nTrackHits method identifies the number of hits per event that fulfill the criteria for the muon track. Fig. \ref{nTrackHits} (left) displays the distribution of such \textit{nTrackHits} for  neutrino-induced cascade-like events (combined MC datasets $\nu^{\rm{atm}}_{e,\mu}$ and $\nu^{\rm{astro}}_{e,\mu}$ shown by blue line), background cascades (red line), and experimental data (black points). Signal cascades have also  non-zero values of \textit{nTrackHits}, which can be caused by mis-identified hits from the noise or cascade. 
	
	Another  method, called BranchRatio, exploits the fact that  for the  $\mu_{\rm{atm}}$ background cascade events  mis-reconstructed as upgoing, more OMs are hit located below the z coordinate of the reconstructed cascade position than above. This method results in the separation variable defined as $\rm{BranchRatio} = \frac{nOMs^{up}}{nOMs^{down}}$.

Neutrino cascades may also be separated from the background using the method referred to as QEarly inspired by the work of the ANTARESS collaboration \cite{ANTARES}. The output of  QEarly is the ratio of  the overall charge of track hits $Q_{\rm{nTrackHits}}$ and total charge of cascade hits $Q {\rm{cascadeHits}}$ (green band in Fig. \ref{trackSketch} (right)), calculated as: 	$\rm{QEarly} = \rm{log_{10}}\bigg(\frac{(Q_{nTrackHits} + a) }{Q_{cascadeHits}}\bigg)$,
where $a = 1$ is a constant that prevents QEarly from reaching infinity.  
		
\subsection{Multivariate Analysis}
After the selection methods have been developed, five output variables from the cascade reconstruction of signal and background simulated datasets were chosen for the multivariate event	classifier BDT within the TMVA package of the CERN ROOT framework \cite{BDT}. Only  cascades reconstructed as upgoing and contained were used in this BDT analysis. The five variables fed into the BDT are: nTrackHits,  BranchRatio, QEarly,  the  Chi-Square after  cascade position reconstruction,  and the reconstructed zenith angle. 
The BDT response score is formed from the five variables given to the TMVA. The BDT response value for signal and background cascades is shown in Fig. \ref{nTrackHits} (right).  The datasets used for the BDT training and testing  have almost the same BDT response curves for both background and signal, however  it can be clearly observed that in the case of background, statistics is the limiting factor. As a result of the BDT analysis relative importance of the individual variables is determined. The QEarly method has the highest separation power, while nTrackHits method is also able to separate between signal and background to a considerable extent.

Subsequently, the optimal cut on the BDT response score (0.48) was determined according to the maximal significance and applied to the experimental data. Signal efficiency is $\approx 49 \%$ after applying the BDT cut and the background has been reduced  to the magnitude on the order of 1 event per cluster per year. Additionaly to the BDT cut, we applied a 50 TeV cut on the reconstructed energy  to search for the high-energy reconstructed cascade events in the experimental data.  One well-reconstructed cascade-like event fulfills the criteria imposed on the BDT output value and reconstructed energy. The event was reconstructed with energy $E$ = 83.3 TeV as contained in Cluster 1 and upgoing direction with zenith angle $\theta$ = 70.9 $\deg$.
For further details on this event refer to Tab. 1. 
	\begin{center}
	\captionof{table}{Reconstructed parameters of the most energetic event found in data 2019 for Cluster1: Cluster, Energy, Zenith angle, Azimuth angle, Distance from the cluster center, Likelihood, Total charge, Total number of hits, Number of hits used in the reconstruction, and Number of track hits.}
	\label{events}
	\begin{tabular}{||c|c| c| c| c| c| c|c|c|c||} 
		\hline
		Cl	&	$E_{\rm{rec}}$ [TeV] & $\theta$ [$\degree$] & $\phi$ [$\degree$] & $\rho$ [m]&L &Q [p.e.]& nHits & nRecoHits & nTrackHits\\ [0.5ex] 
		\hline\hline
		
		1  & 83.3 & 70.9& 4.96 & 47.65   &1.01 &1665.01 &106 & 44 & 1\\
		\hline	
		
	\end{tabular}
\end{center}

	\subsection{Multicluster Search and Preliminary Waveform Comparison}
For the abovementioned upgoing  event, there are some indications that it can  be possibly of the neutrino origin. However, a more complex analysis is required
 to confidently determine if that cascade-like event comes from the neutrino interaction. For that purpose, a search for this cascade event  in the multicluster regime  has been performed. A multicluster event is required to leave a signal in two or more clusters at once.  This detection mode can be very useful as a veto for the background cascades produced along muon tracks. In addition to the detection of background cascade, a long-range muon track  can also be detected in other clusters. Therefore, it is more likely that such event will be found in the multicluster data.
 
\begin{figure}[h!]
	\centering
	\includegraphics[scale=0.5]{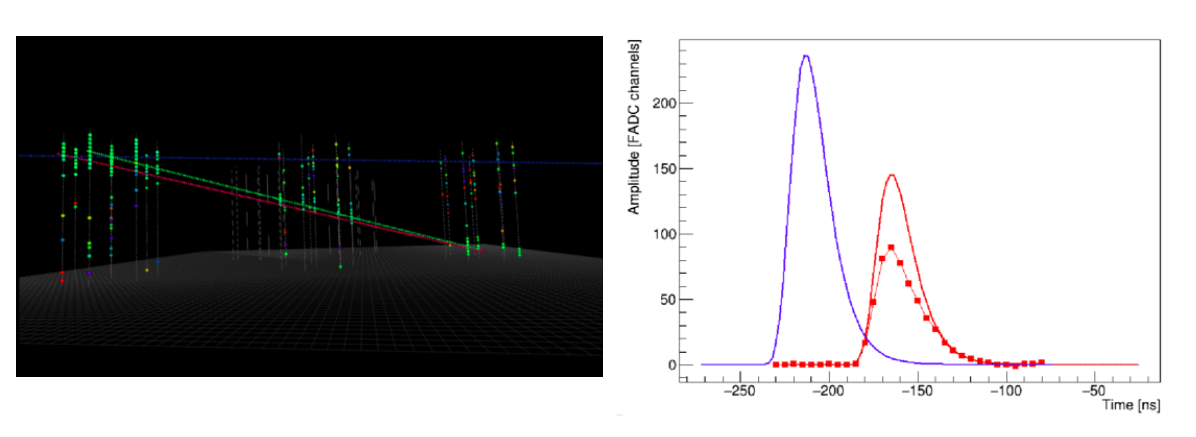}
	\caption{Left: Side view of the multicluster event obtained from the experimental 	data, where the cascade-like event and two muon tracks were independently reconstructed in the single-cluster regime in Cluster 2 (green line), Cluster 3 (blue line). Cascade-like event was reconstructed in Cluster 5 (red line). Right: Waveforms of the pulses from the experimental upgoing event registered at one specific OM are displayed in red with markers. The expected waveform for track hits are displayed in blue. The expected waveform for cascade hits are shown in solid red line. 20 FADC units approximately correspond to 1 p.e.}
	\label{multi}
\end{figure}

 For example, a cascade-like event detected in 2019 on Cluster 5 was also found in multicluster data. This event was reconstructed as downgoing and it passed the standard  reconstruction quality selection criteria.
Indeed,  Fig. \ref{multi} (left) shows that this event was also detected in two other clusters. It suggests that this cascade-like event   comes from the $\mu_{\rm{atm}}$ background, and the corresponding muon was detected in two additional clusters. Subsequently, we tried to search for the  muons independently reconstructed in the single-cluster track reconstruction mode  \cite{Dima}. For this event, two tracks were reconstructed as downgoing, one of them in Cluster2  and the second one
in Cluster3.
We conclude that this
 cascade-like event is most probably of background origin. This procedure was also preformed for the upgoing event described in Tab. 1, in which case the event was not found among multicluster events, suggesting that additional muon tracks are not present. 
 
 After that, the preliminary waveform analysis was performed. In this analysis the
 comparison of the expected waveforms of the pulses (i.e. time and amplitude) induced
 by the cascade or muon track with the real detected waveforms at the OMs was
 performed. The analytical formula for the real pulse waveform can be described by a
 Gumbel function, as follows:
\begin{equation}
		f(t)=A \times e^{-((\frac{t-\mu}{\beta})+e^{-(\frac{t-\mu}{\beta})})},
		\label{Gumbel}
\end{equation}
where $A$ is a scaling amplitude factor, $\mu$ is the time when the pulse reaches maximum amplitude and $\beta$ is the width of Gumbel function. The expected amplitude for the hit coming from the cascade $A^{\rm{cascade}}_i$ is obtained from the pre-calculated MC tables. The expected amplitude for a track hit from a TeV muon can be estimated according to the following function:
\begin{equation}
	A^{\rm{track}}_i \approx e^{-\frac{|\vec{r}_i - \vec{r}_{\rm{track}}|}{\tau}}\cdot \alpha_i,
\end{equation}
where $|\vec{r}_i - \vec{r}_{\rm{track}}|$ is distance between the position of hit OM and  point of light emission from the track (illustrated blue line in Fig. \ref{trackSketch} (left)),
$\tau \approx 24$ m  is the light absorption length in the deep lake water and $\alpha_i$ is the relative sensitivity  of the $i$-th OM as a function of the angle between the OM vertical axis and  incidence of the incoming light.	The  track direction is assumed to be the direction in which the most of track hits were obtained by nTrackHits method.

The  waveform comparison was done for the upgoing  event (Tab. 1). This is illustrated in Fig. \ref{multi} (right),
where the waveform registered by one particular OM is compared to the
cascade and track model predictions. No track hits are observed at the
times. Hence, in accord with the
multicluster analysis, the waveform procedure does not exhibit the presence of muon track for this upgoing event.	
	
\section{Conclusion}
Several selection methods for the neutrino cascades have been developed. They were optimized with Monte Carlo  simulations for season 2019. The corresponding background rejection variables were used for training and testing of a Boosted
Decision Tree (BDT). The BDT was trained and tested with the contained upgoing neutrino and background cascades. Furthermore, experimental data from season 2019 were analyzed.  In the search for the neutrino cascade candidates with higher energy in experimental data we applied cuts on the BDT output value and the reconstructed energy. One contained upgoing event was reconstructed with 83.3 TeV energy. Moreover, a multicluster analysis of that experimental event  and a preliminary waveform comparison were performed. No muon counterparts were found in the multicluster data and the recorded PMT waveforms do not support a muon track origin of the event. Hence this event is likely a genuine neutrino event.

\end{document}